\documentclass[%
 reprint,
 amsmath,amssymb
 aps,
]{revtex4-1}

\usepackage{graphicx}
\usepackage{dcolumn}
\usepackage{bm}
\usepackage{color}
\usepackage{wasysym}
\usepackage{epstopdf}
\usepackage{verbatim}
\usepackage{natbib}
\usepackage{epsfig}
\usepackage{subfigure}
\usepackage{amsmath}
\usepackage{float}

\begin{document}

\preprint{APS/123-QED}

\title{Rapid Quantum Squeezing by Jumping the Harmonic Oscillator Frequency}

\author{Mingjie Xin, Wui Seng Leong, Zilong Chen, Yu Wang}
\author{Shau-Yu Lan}%
 \email{sylan@ntu.edu.sg}
\affiliation{%
Division of Physics and Applied Physics, School of Physical and Mathematical Sciences, Nanyang Technological University, Singapore 637371, Singapore
}%




\date{\today}

\begin{abstract}
Quantum sensing and quantum information processing use quantum advantages such as squeezed states that encode a quantity of interest with higher precision and generate quantum correlations to outperform classical methods. In harmonic oscillators, the rate of generating squeezing is set by a quantum speed limit. Therefore, the degree to which a quantum advantage can be used in practice is limited by the time needed to create the state relative to the rate of unavoidable decoherence. Alternatively, a sudden change of harmonic oscillator's frequency projects a ground state into a squeezed state which can circumvent the time constraint. Here, we create squeezed states of atomic motion by sudden changes of the harmonic oscillation frequency of atoms in an optical lattice. Building on this protocol, we demonstrate rapid quantum amplification of a displacement operator that could be used for detecting motion. Our results can speed up quantum gates and enable quantum sensing and quantum information processing in noisy environments.

\end{abstract}

\pacs{Valid PACS appear here}
\maketitle

The harmonic oscillator is a textbook building block for quantum optics and molecular physics. Its evenly spaced energy levels allow one to construct coherent states that behave in a classical manner, and to assemble nonclassical states. In particular, squeezed states have been used to reduce measurement noise beyond the standard quantum limit by reducing the variance of one quadrature of the oscillator at the expense of increasing the variance of the other quadrature. This is important for quantum metrology and quantum gates in continuous-variable quantum information processing \cite{Deg,Win,Bra,Pez}.

Squeezed states have been prepared in different systems \cite{Win,Pez,Mee,Kie,Lo,Wol,Bur,Man}. In mechanical oscillators, squeezing is typically generated by applying a small perturbative Hamiltonian, which governs quantum evolution from the initial state to the squeezed state. In this case, the rate of creating squeezing is constrained by the quantum speed limit, where the time to generate the state is determined by $\hbar\arccos(|\langle\psi_{\textrm{i}}|\psi_{\textrm{f}}\rangle|)/V$, where $\hbar$ is the reduced Planck constant, $\langle\psi_{\textrm{i}}|\psi_{\textrm{f}}\rangle$ is the overlap of the initial $\psi_{\textrm{i}}$ and final $\psi_{\textrm{f}}$ state wavefunctions, and $V$ is the strength of the perturbation \cite{Man2,Mar,Def}. For instance, motional squeezed states of atoms have been created from the ground state using an external optical dipole force \cite{Mee} and parametric excitation of the trap \cite{Bur} modulated at twice the oscillation frequency. Dissipative reservoir engineering techniques that modulate the potential at the oscillator's frequency have also generated squeezing \cite{Kie,Lo,Wol}. These schemes typically operate at a rate that requires many oscillation periods to reach the available squeezing. Consequently, they can only be applied in oscillators with high quality factor to avoid any associated decoherence faster than the oscillation period. In a noisy oscillator, such as an optical lattice where the potential is not ideally harmonic, squeezing requires free evolution in free space and the operation is not unitary \cite{Mor}.

On the other hand, if the Hamiltonian of a system changes rapidly by an amount of $\Delta \hat{H}=\hat{H}_{2}-\hat{H}_{1}$ over a time $\varepsilon$, the position-space wave function of the particle will remain intact when $|\langle\Delta\hat{H}\rangle|\varepsilon\ll\hbar$ and the wave function can be expressed in a new eigenstates basis. Under this condition, the inverse cosine of the overlap of the initial and final states is zero, so the quantum speed limit does not apply. This could provide rapid quantum state engineering where the decoherence is virtually free. This method has been demonstrated by a noninstantaneous switching of the potential in trapped ions \cite{Wit} and using a thermal state in the optomechanical system \cite{Ras} but has not been realized by a sudden change of a potential in the quantum regime.

\begin{figure}
\includegraphics[scale=0.47]{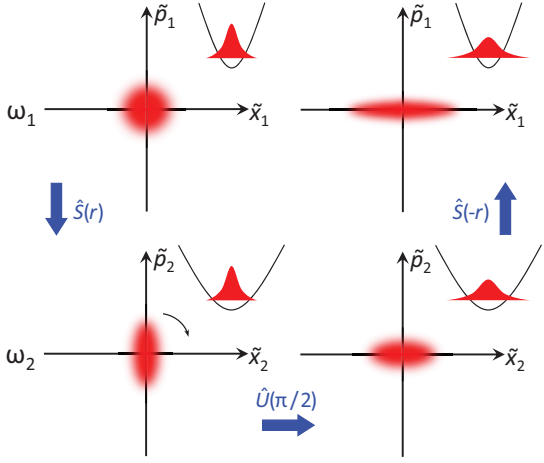}
\caption{\label{fig:Fig1} Illustration of the frequency jump squeezing protocol in a harmonic oscillator in phase and real (inset) spaces. The sequence starts with atoms in the vibrational ground state (upper-left). A sudden jump of the oscillation frequency transforms the atomic wave function into a squeezed state in the new oscillation frequency eigenstates basis (lower-left). After a free oscillation time $\tau_{\textrm{s}}=\pi/2\omega_{2}$, the squeezed state rotates an angle of $\pi/2$ (lower-right) and the oscillation frequency is immediately switched back to $\omega_{1}$ to further squeeze the atomic wave function (upper-right). The resulting squeezing operator is $\hat{S}(-2r)$ for the ground state, and the sequence can continue every quarter period to amplify the squeezing. $\tilde{x}_{i}\equiv x/\sqrt{\hbar/2m\omega_{i}}$ and $\tilde{p}_{i}\equiv p/\sqrt{\hbar m\omega_{i}/2}$ are the canonical position and momentum, where $i=1, 2$.}
\end{figure}

Here, following the textbook model exemplifying nonadiabaticity in a harmonic oscillator and the proposals \cite{Jan,Gra,Sch,Hei,Alo,Alo2} of jumping the oscillator frequency, we control the intensity of the optical lattice laser to generate squeezing and anti-squeezing to atoms in the vibrational ground state. We have achieved 14$^{+3}_{-1}$ dB squeezing from two frequency jumps with one quarter of the oscillation period between the jumps, where each frequency jump generates 7(1) dB squeezing in 1/50 of the oscillation period, three orders of magnitude faster than the previous result of 0.44 dB in one oscillation period \cite{Bur}. Using the same method, we implement a squeezing protocol to amplify a displacement operator created by suddenly shifting the position of the potential minimum.

The harmonic oscillator potential in our system is prepared by an one-dimensional optical lattice. With a beam waist of 100 $\mu$m and 2 W of power per beam at 1064 nm, the vibrational frequency in the axial direction is $\omega_{1}=2\pi\times$93 kHz. Due to anharmonicity, it decreases approximately by 2$\%$ with each increase of vibrational quantum number $n$ with a total of 11 bound states in the lattice potential, see Supplemental Material \cite{SM}. The vibrational frequency is controlled by modulating the lattice beams using acousto-optic modulators (AOMs).

We load an ensemble of cold $^{85}$Rb atoms of 7 $\mu$K from a magneto-optical trap (MOT) into the optical lattice and perform resolved Raman sideband cooling (RSC) on the $|F=2, m_{F}=0\rangle$ state to prepare atoms in the vibrational ground state \cite{Leo}, where $F$ is the hyperfine ground state of $^{85}$Rb and $m_{F}$ is the associated Zeeman state. The mean vibrational quantum number $\bar{n}$ is characterised using vibrational spectroscopy by scanning the two-photon detuning of a pair of counter-propagating Raman beams (797 nm) on the $|F=2, m_{F}=0\rangle$ to the $|F=3, m_{F}=0\rangle$ state. The population in the $|F=3, m_{F}=0\rangle$ state is detected by absorption imaging. We intentionally select a small area of 16$\times$16 $\mu$m$^{2}$ at the trap center for analysis to reduce the broadening of the vibrational frequency due to the radial distribution of atoms. The duration of the Raman pulse is 0.4 ms, much shorter than the radial oscillation period of 4 ms, to ensure atoms are stationary radially.

When the oscillation frequency of a harmonic oscillator is suddenly switched from $\omega_{1}$ to $\omega_{2}$, the Hamiltonian changes from $\hat{H}_{1}=\hat{p}^{2}/2m+m\omega_{1}^{2}\hat{x}^{2}/2=(\hat{a}_{1}^{\dagger}\hat{a}_{1}+1/2)\hbar\omega_{1}$ to $\hat{H}_{2}=\hat{p}^{2}/2m+m\omega_{2}^{2}\hat{x}^{2}/2=(\hat{a}_{2}^{\dagger}\hat{a}_{2}+1/2)\hbar\omega_{2}$ , where $\hat{p}$ is the momentum operator, $\hat{x}$ is the position operator, $\hat{a}_{1}$($\hat{a}_{2}$) and $\hat{a}_{1}^{\dagger}$ ($\hat{a}_{2}^{\dagger}$) are the annihilation and creation operators, and $m$ is the mass. Immediately after the sudden change, the wave function in real space is unaltered while the annihilation and creation operators $\hat{a}_{2}$ and $\hat{a}_{2}^{\dagger}$ undergo a Bogoliubov transformation,

\begin{equation}
\begin{split}
&\hat{a}_{2}=u\hat{a}_{1}-v\hat{a}_{1}^{\dagger} \\
&\hat{a}_{2}^{\dagger}=-v\hat{a}_{1}+u\hat{a}_{1}^{\dagger},
\end{split}
\end{equation}
where $u\equiv\cosh(r)\equiv(\omega_{1}+\omega_{2})/2\sqrt{\omega_{1}\omega_{2}}$, $v\equiv\sinh(r)\equiv(\omega_{1}-\omega_{2})/2\sqrt{\omega_{1}\omega_{2}}$, and $r\equiv\ln(|u+v|)=\ln(\omega_{1}/\omega_{2})/2$. The sudden change of the frequency results in squeezing operation with operator $\hat{S}(\xi)=\exp[(\xi^{\ast}\hat{a}_{1}^{2}-\xi\hat{a}_{1}^{\dagger2})/2]$ along the position quadrature quantified by the squeezing parameter $\xi=r\exp(i2\theta)$ with squeezing amplitude $r$ and angle $\theta=0$, as shown in Fig. 1. Returning the frequency to $\omega_{1}$ after waiting a half-integer number of cycles at $\omega_{2}$ will undo the initial squeezing, while returning it after a quarter-integer number of cycles will increase the squeezing amplitude to 2$r$. The sequence can, in principle, continue for every quarter of the oscillation with a squeezing amplitude $Nr=\ln(\omega_{1}/\omega_{2})^{N}/2$ until the decoherence process dominates, where $N$ is the number of the frequency jumps, see Supplemental Material \cite{SM}.

\begin{figure}
\includegraphics[scale=0.38]{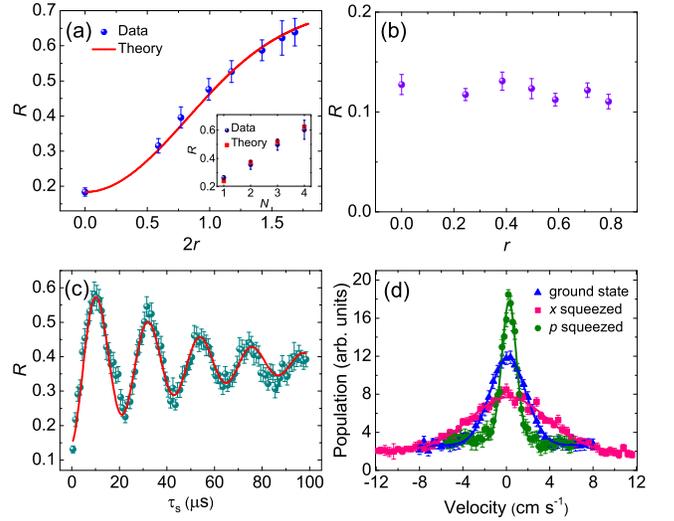}
\caption{\label{fig:Fig2} Frequency jump squeezed state measurements and analysis. (a) Measured $R\equiv P_{-}/P_{+}$ of $\hat{S}(-2r)$ operation versus 2$r$. The red curve is the theoretical model. Inset shows the measured $R$ versus the number of frequency jump $N$ at $r=0.39$. (b) Measured $R$ versus $r$ under $\hat{S}^{\dag}(r)\hat{S}(r)$ operation. (c) Measured $R$ of $\hat{S}^{\dag}(r)\hat{U}(\omega_{2}\tau_{\textrm{s}})\hat{S}(r)$ operation versus the free oscillation time $\tau_{\textrm{s}}$ with 2$r=1.4$. The curve is a fit to the data using exponentially decaying sinusoidal function. (d) Momentum distribution measurements of the ground state, position squeezed state, and momentum squeezed state after releasing the atoms from the potential. The curves are Gaussian functions fitted to the data. The squeezing factor is $e^{2r}=2.58$.}
\end{figure}

We demonstrate squeezing after preparing atoms nearly in their vibrational ground state. The action of applying the sudden switching twice ($N=2$) is a squeezing operator $\hat{S}^{\dag}(r)\hat{U}(\pi/2)\hat{S}(r)=\hat{S}(-2r)$ acting on the ground state $|n=0\rangle$, where the ground state wave function is squeezed along the $p$-quadrature, as shown in Fig. 1. The switching time of the oscillation frequency between $\omega_{1}$ and $\omega_{2}$ is less than 250 ns, much faster than 10.75 $\mu$s oscillation period at $\omega_{1}$, and is limited by the response time of the AOM. To characterise the squeezed state, we adopt the motional sideband analysis method \cite{Mee,Kie,Bur} and measure the vibrational spectrum after the trap returns to $\omega_{1}$. We plot the ratio of the peak population of atoms in the first red sideband $P_{-}$ to that in the first blue sideband $P_{+}$ in the vibrational spectrum $R\equiv P_{-}/P_{+}$ against different 2$r$ through varying $\omega_{2}$ and compare it with the analytical theory, as shown in Fig. 2(a). Our theoretical model includes the imperfect ground state cooling by calculating expectation values of operators for a thermal state density matrix with the initial mean vibrational quantum number $\bar{n}_{0}=0.22$, see Supplemental Material \cite{SM}. The experimental results agree well with the trend of the model of Rabi oscillation for a squeezed thermal state, and the achievable squeezing or 2$r$ is limited by the number of the bound states in our system. The mean vibrational quantum number of a squeezed thermal state and its variance have been calculated \cite{Kim} as $\bar{n}_{st}=\bar{n}_{0}\cosh(4r)+\sinh^{2}(2r)$ and $(\Delta\bar{n}_{\textrm{st}})^{2}=\bar{n}_{0}^{2}\cosh(8r)+\bar{n}_{0}\cosh(8r)+\sinh^{2}(4r)/2$. One standard deviation away from the mean $\bar{n}_{\textrm{st}}+\Delta\bar{n}_{\textrm{st}}$ is close to the 11 bound states in our system, limiting the largest squeezing factor of 10 $\log[e^{4r}]=$ 14$^{+3}_{-1}$ dB that can be attained. The squeezing results with different $N$ at fixed $r=0.39$ is shown in the inset of Fig. 2(a), and its theoretical expectation is plotted.

The unitarity of $\hat{S}(r)$ is confirmed by applying the anti-squeezing operator $\hat{S}^{\dag}(r)$ by jumping the oscillation frequency immediately back to $\omega_{1}$ \cite{Bur}, as shown in Fig. 2(b). No measurable change of $R$ with different $r$ also indicates that the frequency jump squeezing and anti-squeezing are unitary ($\hat{S}^{\dag}(r)\hat{S}(r)=1$), independent of $r$ we introduce here. The coherence of the squeezed state is studied by waiting for the squeezed state in $\omega_{2}$ to evolve for a time $\tau_{\textrm{s}}$ before jumping back to $\omega_{1}$, as shown in Fig. 2(c). The fitted period 21.8(1) $\mu$s of the oscillation is consistent with the expected oscillation frequency $2\omega_{2}=2\pi\times46$ kHz for the squeezed state. The fitted 1/$e$ decay time 46(3) $\mu$s of the measured $R$ is mainly due to the inhomogeneous broadening of the vibrational frequency. We further authenticate our analysis by measuring the atoms’ momentum distribution using Raman velocimetry \cite{Kas}. We compare the distribution of the ground state and the states after the squeezing operations $\hat{S}^{\dag}(r)\hat{U}(\pi/2)\hat{S}(r)=\hat{S}(-2r)$ and $\hat{U}(\pi/2)\hat{S}^{\dag}(r)\hat{U}(\pi/2)\hat{S}(r)=\hat{S}(2r)$ on the ground state, as shown in Fig. 2(d). The measured Gaussian 1/$e^{2}$ velocity width of the ground state is 3.27 cm s$^{-1}$, which is 10$\%$ larger than the calculated quantum fluctuation 2.95 cm s$^{-1}$ at $\omega_{1}$ due to the imperfect ground state. At the squeezing factor of $e^{2r}=2.58$, the momentum width of squeezing in the $p$-quadrature ($\hat{S}(-2r)$) is a factor of 2.43(8) smaller than the width of the ground state, whereas the width of squeezing in the $x$-quadrature ($\hat{S}(2r)$) is a factor of 2.18(8) larger. Our measurements agree with the measurements shown in Fig. 2(a).

\begin{figure}
\includegraphics[scale=0.36]{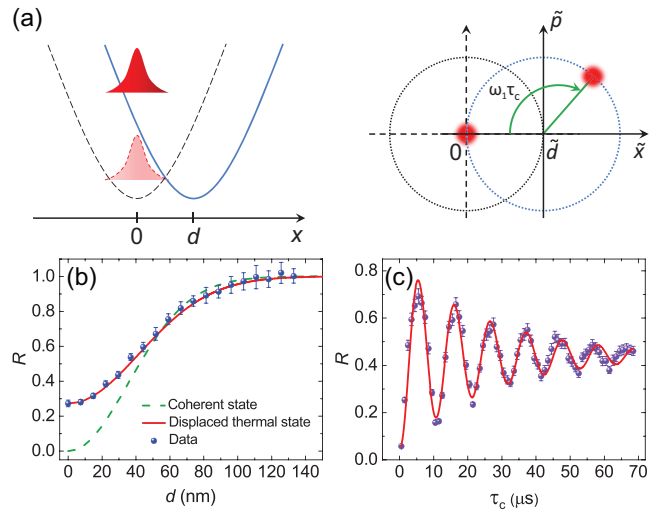}
\caption{\label{fig:Fig3} Measurements and analysis of the coherent state created by a sudden shift of the lattice position. (a) Illustration of the creation of the coherent state, where $\tilde{x}\equiv x/\sqrt{\hbar/2m\omega_{1}}$, $\tilde{d}\equiv d/\sqrt{\hbar/2m\omega_{1}}$, and $\tilde{p}\equiv p/\sqrt{\hbar m\omega_{1}/2}$. Left: real space; right: phase space. (b) Measured $R$ versus the displacement $d$ of the potential minimum. The red curve is the theory of a displaced thermal state and the green dashed curve is the theory for a pure coherent state. (c) Measured $R$ versus the free oscillation time $\tau_{\textrm{c}}$ with $d=29.6$ nm. The curve is a theoretical fit to the data. The fitted 1/$e$ decay time is 27(2) $\mu$s.}
\end{figure}

Quantum squeezing has been investigated as a valuable tool in amplifying interaction strength to speed up quantum dynamics for alleviating decoherence \cite{Bur,Bur2,Are}. It has been shown that squeezing operators can amplify a coherent state through parametric modulation on the potential \cite{Bur,Bur2}. The ability to speed up the amplification process could allow for filtering out any unwanted noise during amplification. Here, we amplify a coherent state created by rapidly shifting the minimum of the trapping potential.

The coherent state $|\alpha\rangle$ can be created by applying a displacement operator $\hat{D}(\alpha)=\exp(\alpha\hat{a}^{\dag}-\alpha^{\ast}\hat{\alpha})$ on the ground state $|n=0\rangle$, where $\hat{a}$ and $\hat{a}^{\dag}$ are the corresponding annihilation and creation operators, and $\alpha=|\alpha|\exp(i\varphi)$. The displacement operator has been realised experimentally in mechanical oscillators using resonant driving through modulating the trapping potential or an external optical dipole force whose interaction Hamiltonian mimics the displacement operator \cite{Mee,Mor,Leo,Joh}. Another way is to suddenly translate the minimum of the oscillator potential \cite{Alo2,Lam} by a distance $d$, as shown on the left of Fig. 3(a). The action is a translation operator $\hat{T}\equiv\exp(-i\hat{p}d/\hbar)=\exp(d(\hat{a}^{\dag}-\hat{a})/2x_{0})$, where $\hat{p}=i\hbar(\hat{a}^{\dag}-\hat{a})/2x_{0}$. The coherent state is then related to the displacement as $\alpha=d/2x_{0}$, where $x_{0}=\sqrt{\hbar/2m\omega}$ is the root-mean-square extent of the ground state wave function and $\varphi=0$.

\begin{figure}
\includegraphics[scale=0.36]{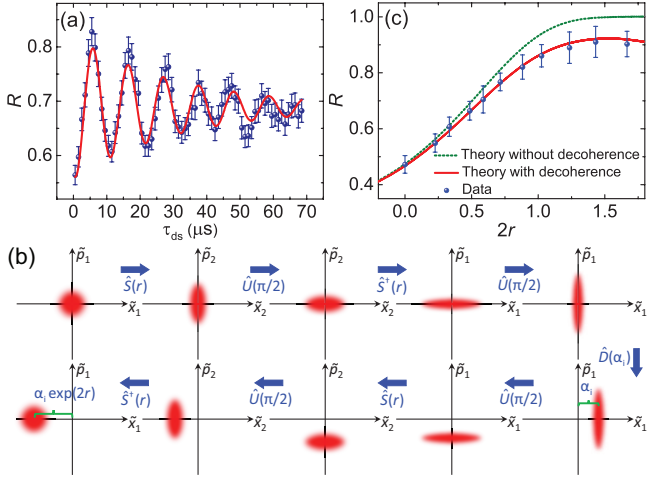}
\caption{\label{fig:Fig4} Measurements and analysis of the amplified coherent state by the frequency jump squeezing. (a) Measured $R$ versus the free oscillation time $\tau_{\textrm{ds}}$ of the displaced squeezed state under $\hat{D}^{\dagger}(\alpha_{\textrm{i}})\hat{U}(\omega_{1}\tau_{\textrm{ds}})\hat{D}(\alpha_{\textrm{i}})\hat{U}(\pi/2)\hat{S}^{\dag}(r)\hat{U}(\pi/2)\hat{S}(r)$ operation with $\alpha_{\textrm{i}}=0.67$ and $2r=1.23$. The curve fits the data using exponentially decaying sinusoidal function. The fitted 1/$e$ decay time $\Gamma=32(4)$ $\mu$s. (b) Phase space illustration of the amplification protocol sequence. (c) Measured $R$ versus the squeezing parameter $2r$ using the amplification operation shown in (b). The red curve is the theory including the decoherence, whereas the green dashed curve is the one without decoherence.}
\end{figure}

We jump the position of the optical lattice by changing the phase of one of the lattice lasers controlled by the radio wave fed into the AOM. The size of a coherent state is then determined from measurements using vibrational spectroscopy. Figure 3(b) shows the measured $R$ as a function of the shift $d$ of the lattice position. At $d=133$ nm, the obtained $\alpha=3$ and its $\bar{n}=\alpha^{2}=9$ is close to the available bound states in the system. We confirm our measurements by the theoretical model using displaced thermal state, see Supplemental Material \cite{SM}. For comparison, we also plot the curve for a pure coherent state. Our result starts with thermal-state-like and approaches a coherent state for large $d$.

After a free oscillation time $\tau_{\textrm{c}}$ in the displaced potential, we undo the phase shift, letting the potential minimum return to the original position, as shown on the right of Fig. 3(a). The new coherent state oscillates as a function of the time $\tau_{\textrm{c}}$ with frequency $\omega_{1}$. Figure 3(c) shows the measured $R$ as a function of $\tau_{\textrm{c}}$. We fit the data with exponentially decaying sinusoidal function, and the fitted oscillation period 10.5(1) $\mu$s matches with the oscillation frequency $\omega_{1}=2\pi\times93$ kHz.

For coherent state amplification, we first create the displaced squeezed state by applying $\hat{D}(\alpha_{\textrm{i}})\hat{U}(\pi/2)\hat{S}^{\dag}(r)\hat{U}(\pi/2)\hat{S}(r)$ to atoms in the ground state using our frequency jump protocol, where $\alpha_{\textrm{i}}$ is the initial displacement. The state is verified by applying the operator $\hat{D}^{\dag}(\alpha_{\textrm{i}})$ after a free oscillation time $\tau_{\textrm{ds}}$. As shown in Fig. 4(a), the measured $R$ oscillates with the period 10.6(1) $\mu$s, matching $\omega_{1}$ instead of $2\omega_{1}$ as in the squeezed vacuum state. Similar to the protocol in Ref. \cite{Bur}, we apply the operator $\hat{S}^{\dag}(r)\hat{U}(\pi/2)\hat{S}(r)$ to anti-squeeze the displaced squeezed state after a free oscillation time $\pi/2\omega_{1}$. Using the identity $\hat{D}(\alpha_{\textrm{i}})\hat{S}(r)=\hat{S}(r)\hat{D}(\alpha_{\textrm{i}}\exp(r))$, where a displaced squeezed state is different from a squeezed coherent state \cite{Man}, the resulting operator is another displaced operator

\begin{equation}
\begin{split}
\hat{D}(\alpha_{\textrm{f}})=[\hat{S}^{\dag}(r)\hat{U}(\pi/2)\hat{S}(r)\hat{U}(\pi/2)] \\
\otimes\hat{D}(\alpha_{\textrm{i}})[\hat{U}(\pi/2)\hat{S}^{\dag}(r)\hat{U}(\pi/2)\hat{S}(r)].
\end{split}
\end{equation}

The initial coherent state is, thus, amplified to the final state $\alpha_{\textrm{f}}=\alpha_{\textrm{i}}\exp(2r)\exp(i\pi)$ with a $\pi$ phase shift as shown in Fig. 4(b). Figure 4(c) shows the measured $R$ as a function of the squeezing parameter 2$r$ with the same initial coherent state $\alpha_{\textrm{i}}=0.67$. The red curve is the theory that considers the imperfect ground state and the decoherence during the free oscillation time, see Supplemental Material \cite{SM}. The amplification process only requires the wave packet to oscillate for one cycle, as opposed to modulation methods \cite{Bur,Bur2}.

We experimentally realize unitary quantum operations based on sudden changes in the oscillation frequency of a harmonic oscillator, including squeezing, displacement, and amplification, which is intrinsically resilient to environmental noises. We demonstrate such resilience in a noisy oscillator. The quality factor of our oscillator is approximately only 5, in which the state decays even with just one cycle of the oscillation. It is then challenging to create a squeezed state in the presence of large decoherence using the modulation methods that require many cycles of oscillation. The largest squeezing in our experiment is limited by the available bound states and not the decoherence nor interaction strength. Compared to other quantum squeezing methods, this frequency jump protocol does not reply on quantum evolution and can be instantaneous in any frame of reference. This method is universal to any quantum harmonic oscillator that can jump its oscillation frequency faster than its oscillation period and can be useful in speeding up the entanglement generation and quantum gates operation in continuous-variable quantum information \cite{Bra}. The rapid quantum state engineering in a harmonic oscillator presented in this work can also offer new opportunities in trapped atom interferometers using nonclassical states for sensing \cite{Lo,Joh2,Cam,Shi,Sut}.

We thank Dzmitry Matsukevich and Holger M\"{u}ller for conversations and Shijie Chai for the assistance on the dipole laser. This work was financially supported by Singapore National Research Foundation under Grant No. QEP-P4, Singapore Ministry of Education under Grant No. MOE2017-T2-2-066 and RG80/20.


\nocite{*}

\bibliography{apssamp}

\begin{thebibliography}{00}
\bibitem{Deg}C.L. Degen, F. Reinhard, and P. Cappellaro, Quantum sensing. Rev. Mod. Phys. {\bf89}, 035002 (2017).
\bibitem{Win}D. J. Wineland, Nobel lecture. Superposition, entanglement, and raising Schr\"{o}dinger’s cat. Rev. Mod. Phys. {\bf85}, 1103 (2013).
\bibitem{Bra}S. L. Braunstein and P. van Loock, Quantum information with continuous variables. Rev. Mod. Phys. {\bf77}, 513 (2005).
\bibitem{Pez}L. Pezz\`{e}, A. Smerzi, M. K. Oberthaler, R. Schmied, and P. Treutlein, Quantum metrology with nonclassical states of atomic ensembles. Rev. Mod. Phys. {\bf90}, 035005 (2018).
\bibitem{Mee}D. M. Meekhof, C. Monroe, B. E. King, W. M. Itano, and D. J. Wineland, Generation of nonclassical motional states of a trapped atom. Phys. Rev. Lett. {\bf76}, 1796 (1996).
\bibitem{Kie}D. Kienzler, H.-Y. Lo, B. Keitch, L. de Clercq, F. Leupold, F. Lindenfelser, M. Marinelli, V. Negnevitsky, and J. P. Home, Quantum harmonic oscillator state synthesis by reservoir engineering. Science {\bf347}, 53 (2015).
\bibitem{Lo}H.-Y. Lo, D. Kienzler, L. de Clercq, M. Marinelli, V. Negnevitsky, B. C. Keitch, and J. P. Home, Spin-motion entanglement and state diagnosis with squeezed oscillator wavepackets. Nature {\bf521}, 336 (2015).
\bibitem{Wol}E. E. Wollman, C. U. Lei, A. J. Weinstein, J. Suh, A. Kronwald, F. Marquardt, A. A. Clerk, and K. C. Schwab, Quantum squeezing of motion in a mechanical resonator. Science {\bf349}, 952 (2015).
\bibitem{Bur}S. C. Burd, R. Srinivas, J. J. Bollinger, A. C. Wilson, D. J. Wineland, D. Leibfried, D. H. Slichter, and D. T. C. Allcock, Quantum amplification of mechanical oscillator motion. Science {\bf364}, 1163 (2019).
\bibitem{Man}L. Mandel and E. Wolf, \textit{Optical Coherence and Quantum Optics} (Cambridge Univ. Press, 1995).
\bibitem{Man2}L. Mandelstam and I. G. Tamm, The uncertainty relation between energy and time in non-relativistic quantum mechanics. J. Phys. USSR {\bf9}, 249 (1945).
\bibitem{Mar}N. Margolus and L. B. Levitin, The maximum speed of dynamical evolution. Physica. D {\bf120}, 188 (1998).
\bibitem{Def}S. Deffner and S. Campbell, Quantum speed limits: from Heisenberg uncertainty principle to optimal quantum control. J. Phys. A: Math. Theor. {\bf50}, 453001 (2017).
\bibitem{Mor}M. Morinaga, I. Bouchoule, J. C. Karam, and C. Salomon, Manipulation of motional quantum states of neutral atoms. Phys. Rev. Lett. {\bf83}, 4037 (1999).
\bibitem{Wit}M.Wittemer, F. Hakelberg, P. Kiefer, J.-P. Schr\"{o}der, C. Fey, R. Schützhold, U. Warring, and T. Schaetz, Phonon Pair Creation by Inflating Quantum Fluctuations in an Ion Trap, Phys. Rev. Lett. {\bf123}, 180502 (2019).
\bibitem{Ras}M. Rashid, T. Tufarelli, J. Bateman, J. Vovrosh, D. Hempston, M. S. Kim, and H. Ulbricht, Experimental realisation of a thermal squeezed state of levitated optomechanics. Phys. Rev. Lett. {\bf117}, 273601 (2016).
\bibitem{Jan}J. Janszky and Y. Y. Yushin, Squeezing via frequency jump. Optics Communications, {\bf59}, 151 (1986).
\bibitem{Gra}R. Graham, Squeezing and frequency changes in harmonic oscillations. Journal of Modern Optics, {\bf34}, 873 (1987).
\bibitem{Sch}W. Schleich and J. Wheeler, Oscillations in photon distribution of squeezed states and interference in phase space. Nature {\bf326}, 574 (1987).
\bibitem{Hei}D. J. Heinzen and D. J. Wineland, Quantum-limited cooling and detection of radio-frequency oscillations by laser-cooled ions. Phys. Rev. A {\bf42}, 2977 (1990).
\bibitem{Alo}J. Alonso, F. M. Leupold, B. C. Keitch, and J. P. Home, Quantum control of the motional states of trapped ions through fast switching of trapping potentials. New J. Phys. {\bf15}, 023001 (2013).
\bibitem{Alo2}J. Alonso, F. M. Leupold, Z. U., Sol\`{e}r, M. Fadel, M. Marinelli, B. C. Keitch, V. Negnevitsky, and J. P. Home, Generation of large coherent states by bang–bang control of a trapped-ion oscillator. Nat Commun {\bf7}, 11243 (2016).
\bibitem{SM}see Supplemental Material.
\bibitem{Leo}W. S. Leong, M. Xin, Z. Chen, S. Chai, Y. Wang, and S.-Y. Lan, Large array of Schr\"{o}dinger cat states facilitated by an optical waveguide. Nat Commun {\bf11}, 5295 (2020).
\bibitem{Kim}M. S. Kim, F. A. M. de Oliveira, and P. L. Knight, Properties of squeezed number states and squeezed thermal states. Phys. Rev. A {\bf40}, 2494 (1989).
\bibitem{Kas}M. Kasevich, D. S. Weiss, E. Riis, K. Moler, S. Kasapi, and S. Chu, Atomic velocity selection using stimulated Raman transitions. Phys. Rev. Lett. {\bf66}, 2297 (1991).
\bibitem{Bur2}S. C. Burd, R. Srinivas, H. M. Knaack, W. Ge, A. C. Wilson, D. J. Wineland, D. Leibfried, J. J. Bollinger, D. T. C. Allcock, and D. H. Slichter, Quantum amplification of boson-mediated interactions. Nat. Phys. {\bf17}, 898 (2021).
\bibitem{Are}C. Arenz, D. I. Bondar, D. Burgarth, C. Cormick, and H. Rabitz, Amplification of quadratic Hamiltonians. Quantum {\bf4}, 271 (2020).
\bibitem{Joh}K. G. Johnson, J. D. Wong-Campos, B. Neyenhuis, J. Mizrahi, and C. Monroe, Ultrafast creation of large Schr\"{o}dinger cat states of an atom. Nat Commun {\bf8}, 697 (2017).
\bibitem{Lam}M. R. Lam, N. Peter, T. Groh, W. Alt, C. Robens, D. Meschede, A. Negretti, S. Montangero, T. Calarco, and A. Alberti, Demonstration of quantum brachistochrones between distant states of an atom. Phys. Rev. X {\bf11}, 011035 (2021).
\bibitem{Joh2}K. G. Johnson, B. Neyenhuis, J. Mizrahi, J. D. Wong-Campos, and C. Monroe, Sensing Atomic motion from the zero point to room temperature with ultrafast atom interferometry. Phys. Rev. Lett. {\bf115}, 213001 (2015).
\bibitem{Cam}W. C. Campbell and P. Hamilton, Rotation sensing with trapped ions. J. Phys. B {\bf50}, 064002 (2017).
\bibitem{Shi}A. Shinjo, M. Baba, K. Higashiyama, R. Saito, and T. Mukaiyama, Three-dimensional matter-wave interferometry of a trapped single ion. Phys. Rev. Lett. {\bf126}, 153604 (2021).
\bibitem{Sut}R. T. Sutherland, S. C. Burd, D. H. Slichter, S. B. Libby, and D. Leibfried,Motional Squeezing for Trapped Ion Transport and Separation, Phys. Rev. Lett. {\bf127}, 083201 (2021).
\bibitem{Jan2}J. Janszky and P. Adam, Strong squeezing by repeated frequency jumps. Phys. Rev. A, {\bf46}, 6091 (1992).
\bibitem{Oli}F. A. M. de Oliveira, M. S. Kim, P. L. Knight, and V. Bu\v{z}ek, Properties of displaced number states. Phys. Rev. A {\bf41}, 2645 (1990).
\end{thebibliography}

\maketitle
\section*{Supplemental Material}
\section{Anharmonicity in the optical lattice potential}
The atomic wave function in the optical lattice of potential $V_{0}$ is governed by the stationary Schrödinger equation, identical to the Mathieu equation, as

\begin{equation}
\frac{\partial^{2}\Phi_{n}(\chi)}{\partial\chi^{2}}+[a_{n}-2q\cos(2\chi)]\Phi_{n}(\chi)=0,
\end{equation}
where $\Phi_{n}(\chi)$ is the wave function of order $n$, $\chi\equiv kx$, $x$ is the spatial coordinate, $k$ is the wavenumber of the lattice beams, $a_{n}\equiv E_{n}/E_{\textrm{R}}-2q$, $E$ is the eigenenergy, $E_{\textrm{R}}=\hbar^{2}k^{2}/2m$ is the recoil energy, and $q\equiv V_{0}/4E_{\textrm{R}}$. In the limit of $q\gg1$, the asymptotic expansion of eigenenergy can be written as

\begin{equation} \label{anhar}
\begin{split}
\frac{E_{n}}{E_{\textrm{R}}}=2(2n+1)\sqrt{q}-\frac{(2n+1)^{2}+1}{2^{3}} \\
-\frac{(2n+1)^{3}+3(2n+1)}{2^{7}\sqrt{q}}-\cdots.
\end{split}
\end{equation}

For the maximum potential of $V_{0}=\hbar \times 2\pi \times 1.05$ MHz and $E_{\textrm{R}}=\hbar \times 2\pi \times 2$ kHz of 1064 nm lattice beam wavelength in our experiment, the available energy levels in our system are $n=11$. Considering only the first two leading terms in equation (\ref{anhar}), the energy gaps between states are
\begin{equation}
E_{n+1}-E_{n}=\hbar\omega-(n+1)E_{\textrm{R}}.
\end{equation}
The vibrational frequency, therefore, decreases by one recoil frequency with every increase of $n$, where $\omega=4\sqrt{q}E_{\textrm{R}}/\hbar$ is the oscillation frequency of the approximated harmonic potential.

\section{Squeezing by multiple frequency jumps}
The squeezing can be increased or decreased by waiting for some evolution time. This can be seen from the annihilation operator $\hat{a}_{1}'=u'\hat{a}_{1}+v'\hat{a}_{1}^{\dagger}$ and creation operator $\hat{a}_{1}'^{\dagger}=u'^{\ast}\hat{a}_{1}^{\dagger}+v'^{\ast}\hat{a}_{1}$ in the old ($\omega_{1}$) trap after the second frequency jump with a waiting time $\tau_{\textrm{s}}$, where $u'=\cos(\omega_{2}\tau_{\textrm{s}})-i(u^{2}+v^{2})\sin(\omega_{2}\tau_{\textrm{s}})$ and $v'=-2iuv\sin(\omega_{2}\tau_{\textrm{s}})$ [35]. At $\omega_{2}\tau_{\textrm{s}}=s\pi$, where $s=0, 1, 2\ldots$, an anti-squeezing operator $\hat{S}(-r)=\hat{S}^{\dagger}(r)$ is created along the position quadrature to reverse the first squeezing operation. At $\omega_{2}\tau_{\textrm{s}}=(s+1/2)\pi$, the squeezing amplitude is increased to $\ln(|u'+v'|)=\ln(\omega_{1}/\omega_{2})^{2}/2=2r$ with the resulting operator $\hat{S}^{\dag}(r)\hat{U}((s+1/2)\pi)\hat{S}(r)=\hat{S}(-2r)$ for a ground state. Here, $\hat{U}((s+1/2)\pi)$ is the free oscillation operator on the wave packet for $(s+1/2)\pi$ radian. The sequence can continue for every quarter of the oscillation with a squeezing amplitude $Nr=\ln(\omega_{1}/\omega_{2})^{N}/2$, where $N$ is the number of the frequency jumps.

\begin{figure}
\includegraphics[scale=0.3]{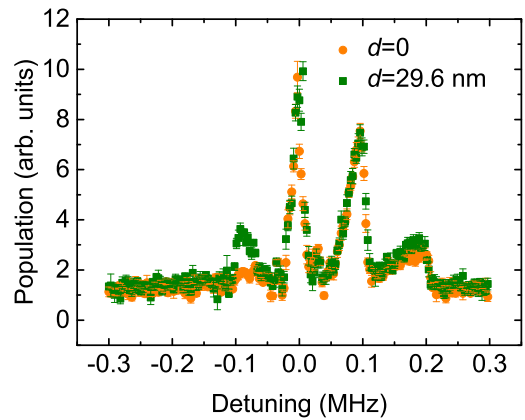}
\caption{\label{fig:EDFig1} Vibrational spectrum. The population of atoms in $|F=3, m_{F}=0\rangle$ is detected through absorption imaging with different frequency detuning between two Raman beams. The orange circles are the spectrum after RSC, where the mean vibrational number after RSC is defined as $\bar{n}_{0}=R/(1-R)$. The green squares are the spectrum after a sudden displacement of the lattice position by $d=29.6$ nm. The peak values of the first blue sideband and red sideband are used to define $R$.}
\end{figure}
\begin{figure}
\includegraphics[scale=0.3]{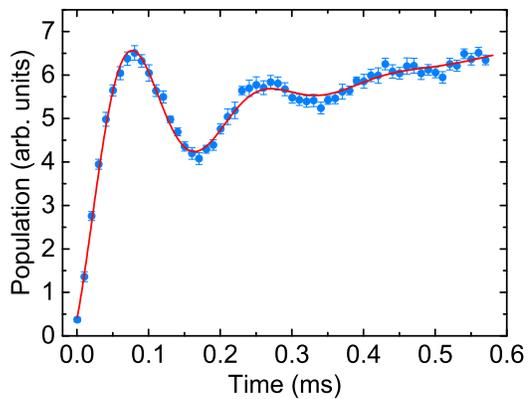}
\caption{\label{fig:EDFig2} Rabi flopping of $|F=2, m_{F}=0, n=0\rangle$ to $|F=3, m_{F}=0, n=1\rangle$. The red curve fits the data using the function $A+B\exp(-\gamma t)\sin(\Omega_{0,1}t+\Theta)+C(1-\exp(-t/t_{2}))$, where the first term is the background, the second term is the Rabi oscillation, and the third term includes the slow drift of the population.}
\end{figure}
\section{Squeezed state and coherent state analysis}

We characterise the quantum states by analysing their vibrational state population distribution. This is done by taking the ratio $R$ of the first red sideband peak population to the first blue sideband peak population in the vibrational spectrum, as shown in Fig. ~\ref{fig:EDFig1}. The first blue ($P_{+}$) or first red ($P_{-}$) sideband population is measured when the relative frequency of the Raman beams is blue or red detuned at the vibrational frequency. The population of each sideband after a Raman pulse duration $t$ is proportional to the sum of the probability in each vibrational state after Rabi flopping as

\begin{equation}
\begin{split}
&P_{+}(t)\propto \frac{1}{2}\sum^{\infty}_{n=0}P_{n}(1-e^{-\gamma t}\cos(\sqrt{n+1}\Omega_{0,1}t))\\
&P_{-}(t)\propto \frac{1}{2}\sum^{\infty}_{n=1}P_{n}(1-e^{-\gamma t}\cos(\sqrt{n}\Omega_{0,1}t)),
\end{split}
\end{equation}
where $P_{n}$ is the probability in the state $|n\rangle$, $\Omega_{0,1}$ is the two-photon Rabi frequency of $|F=2, m_{F}=0, n=0\rangle$ to $|F=3, m_{F}=0, n=1\rangle$ transition, and $\gamma$ is the decay rate of the Rabi flopping. We calculate the expected population $P_{\pm}(t)$ and compare it with our measurements. In the calculation, $t=0.4$ ms, and $\Omega_{0,1}=2\pi\times5.4$ kHz and $\gamma=9.8$ kHz are determined from the measured Rabi flopping, as shown in Fig. ~\ref{fig:EDFig2}. The infinite sum is replaced by the sum to $n=20$ in our calculation.

The probability $P_{n}$ takes into account the imperfect RSC by weighing the Boltzmann distribution as

\begin{equation} \label{Pro}
P_{n}=\sum^{\infty}_{l=0}\frac{1}{1+\bar{n}_{0}}\left(\frac{\bar{n}_{0}}{1+\bar{n}_{0}}\right)^{l}|\langle n|\hat{M}|l\rangle|^{2},
\end{equation}
where $\bar{n}_{0}$ is the initial mean vibrational number, and $\hat{M}$ is the operator of interest. For a squeezing operator $\hat{S}(r)$, the matrix element has been calculated as [25]

\begin{equation}
\begin{split}
|\langle n|\hat{S}(r)|l\rangle|^{2}=\frac{l!n!}{(\cosh(r))^{2n+1}}\left(\frac{\tanh(r)}{2}\right)^{l-n} \\
\times\left|\sum^{n/2}_{g=\frac{n-l}{2}}\frac{(-1)^{g}(\frac{\sinh(r)}{2})^{2g}}{g!(n-2g)![\frac{(l-n)}{2}+g]!}\right|^{2}.
\end{split}
\end{equation}
For a coherent operator $\hat{D}(\alpha)$, the matrix element has been calculated as [36]

\begin{equation}
\begin{split}
&|\langle n|\hat{D}(\alpha)|l\rangle|^{2}=\frac{\exp(-|\alpha|^{2})}{l!n!}|\alpha|^{2|l-n|} \\
&\times\left|\sum^{\min(l,n)}_{g=0}\binom{l}{g}\binom{n}{g}g!(-|\alpha|^{2})^{\min(l,n)-g}\right|^{2}.
\end{split}
\end{equation}
The red curve in Fig. 3(b) is the theoretical model based on the experimental parameters $\bar{n}_{0}=0.38$, $\Omega_{0,1}=2\pi\times5.4$ kHz, and $\gamma=9.8$ kHz and the pure coherent state is obtained by taking $\bar{n}_{0}=0$ of equation (\ref{Pro}).\newline

\section{Coherent state amplification analysis}

The amplification protocol involves four squeezing operations, one displacement operation, and four quarters of free oscillation. We assume that the decoherence in the amplification protocol leads to the increase of thermal noise and is mainly due to dephasing during the free oscillation time. We, therefore, weigh the vibrational state probability after the operation with a phenomenological time-dependent exponential function as

\begin{align}
\begin{split}
P_{n}(t')=&\exp(\frac{-t'}{\Gamma})\sum^{\infty}_{l=0}\frac{1}{1+\bar{n}_{0}}\left(\frac{\bar{n}_{0}}{1+\bar{n}_{0}}\right)^{l}|\langle n|\hat{D}(\alpha_{\textrm{f}})|l\rangle|^{2} \\
&+\frac{1-\exp(\frac{-t'}{\Gamma})}{1+\bar{n}_{0}+|\alpha_{\textrm{f}}|^{2}}\left(\frac{\bar{n}_{0}+|\alpha_{\textrm{f}}|^{2}}{1+\bar{n}_{0}+|\alpha_{\textrm{f}}|^{2}}\right)^{n},
\end{split}
\end{align}
where $\Gamma=32$ $\mu$s is the decay time obtained from the fitted value in Fig. 4(a). The first term is the amplified coherent state contribution, and the second term is the thermal state contribution. When $t'=0$, the probability distribution is for an amplified displaced thermal state. When $t'$ approaches infinity, the distribution is thermalized with $\bar{n}=\alpha_{\textrm{f}}^{2}$. The total free oscillation time in our experiment is $t'=\pi/\omega_{1}+\pi/\omega_{2}$ and the theoretical model also considers the imperfect ground state $\bar{n}_{0}=0.35$.

\end{document}